\documentstyle[prl,aps,multicol,epsf]{revtex}
\begin{document}
\draft
\title{Scaling of the magnetoconductivity of 
silicon MOSFET's:\\evidence for a quantum phase transition in two dimensions.}
\author{S.~A.~Vitkalov, H. Zheng, K. M. Mertes and M.~P.~Sarachik}
\address{Physics Department, City College of the City
University of New York, New York, New York 10031}
\author{T.~M.~Klapwijk}
\address{Delft University of Technology, Department of Applied Physics,
2628 CJ Delft, The Netherlands}
\date{\today}
\maketitle

\begin{abstract}
For a broad range of electron densities $n$ and temperatures $T$, the in-plane
magnetoconductivity of the two-dimensional system of electrons in silicon
MOSFET's can be scaled onto a universal curve with a single parameter
$H_{\sigma}(n,T)$, where $H_{\sigma}$ obeys the empirical relation $H_{\sigma}=A
(n) [\Delta(n)^2 +T^2]^{1/2}$.  The characteristic energy $k_B \Delta$ associated
with the magnetic field dependence of the conductivity decreases with decreasing
density, and extrapolates to $0$ at a critical density $n_0$, signaling the
approach to a zero-temperature quantum phase transition.  We show that
$H_{\sigma}=AT$ for densities near $n_0$.

\end{abstract}


\begin{multicols}{2}

Strongly interacting two-dimensional systems of electrons (or holes) have drawn
intensive recent attention \cite{Abrahams} due to their anomalous behavior: their
resistance exhibits metallic temperature-dependence above a critical electron
\cite{krav,popovic} (or hole \cite{coleridge,shahar,cambridge}) density $n_c$,
raising the possibility of an unexpected metallic phase in two dimensions.  An
equally intriguing characteristic of these systems is their enormous response to
magnetic fields applied in the plane of the electrons: the resistance increases 
dramatically with in-plane magnetic field and saturates to a new value above a
characteristic magnetic field $H_{sat}$ on the order of several tesla
\cite{dolgopolov,simonian,pudalov,yoon}.  For high electron densities,
measurements of Shubnikov-de Haas oscillations
\cite{okamoto,vitkalov,vitkalovangular} have established that the magnetic field
$H_{sat}$ is the field at which full polarization of the electrons is reached. 
A parallel magnetic field has been shown to suppress the metallic
temperature-dependence \cite{simonian,Kevin}.  Data obtained by Pudalov {\it et
al.} \cite{Pudalov2} and by Shashkin {\it et al} \cite{Shashkin} indicate there
is a substantial increase in the g-factor as the electron density is decreased
toward $n_c$.  These experimental findings all suggest that the behavior of
the spins is key to understanding the enigmatic behavior of dilute, strongly
interacting systems in two dimensions.

In this paper we report measurements of the temperature-dependence and
density-dependence of the in-plane magnetoconductivity of silicon
metal-oxide-semiconductor field effect transistors (MOSFET's).  For a broad
range of electron densities and temperatures, we show that all data for the
magnetoconductance can be collapsed onto a single universal curve using a
single parameter
$H_{\sigma}$ which obeys an empirical relation given by $H_{\sigma}(n,T)=A (n)
[\Delta(n)^2 +T^2]^{1/2}$.  The characteristic energy $k_B\Delta$
associated with the response to magnetic field is found to decrease with
decreasing electron density, and to exhibit critical behavior, extrapolating to
$0$ at a density $n_0$ near the critical density $n_c$ for the zero-field
metal-insulator transition.  $H_{\sigma}=AT$ for densities near $n_0$, 
so that the magnetoconductivity scales with $H/T$.  Our results provide strong
experimental evidence for a zero-temperature quantum phase transition at density
$n_0$.  We suggest that this is a transition to a ferromagnetically ordered
state in two dimensions. 

Measurements were taken on three silicon MOSFETs: the mobility $\mu \approx
30,000\;$V/(cm$^2s)$ at $0.3$ K for sample \#$1$ and $\approx
20,000\;$V/(cm$^2s)$ for samples \#$2$ and \#$3$.  Data were obtained on samples
with split-gate geometry at City College to $12$ T and at the National Magnetic
Field Laboratory in fields up to $20$ T using standard four-terminal AC
techniques described elsewhere
\cite{vitkalov}.

Fig. 1 (a) shows the longitudinal conductivity $\sigma_{xx}$ as a function
of magnetic field $H_{||}$ applied parallel to the plane of a silicon
MOSFET for different electron densities $n > n_c \approx 0.85 \times
10^{11}$ cm$^{-2}$ ($n_c$ is the critical density for the zero-field
metal-insulator transition).  In agreement with earlier results, the
conductivity decreases (resistivity increases) dramatically with increasing
magnetic field and saturates to a value $ \sigma (H=\infty)$ that is almost
independent of magnetic field in fields $H >H_{\sigma}$.

Attempts \cite{yoon,Pudalov2,Shashkin,Dolgopol2} to obtain a collapse of the
magnetoresistance onto a single scaled curve have generally yielded scaling at
either low or high magnetic field, but not over the entire field range.  We now
report a method that allows a full mapping of the magnetoconductivity (rather
than the magnetoresistivity) onto a single universal curve over the entire range
of magnetic field for all data obtained at different densities $n$ and
temperatures $T$, as follows.  We separate the conductivity into a
field-dependent contribution $[\sigma - \sigma (H=\infty)]$, which was shown
\cite{okamoto,vitkalov} to reflect the degree of spin polarization of the
electrons, and a contribution that is independent of magnetic field,
$\sigma (H=\infty)$.  We find that the field-dependent contribution to the
conductivity, $\sigma(H=0)-\sigma(H)$, normalized to its full value,
$\sigma(H=0)-\sigma(H=\infty)$ is a universal function of $H/H_{\sigma}$
\cite{Altshuler}, so that: 

$$\frac{\sigma(H=0)-\sigma(H)}{\sigma(H=0)-\sigma(H=\infty)}=F(H/H_{\sigma})
\eqno{(1)}
$$
using a single scaling parameter $H_{\sigma}(n,T)$  which is a function of
electron density $n$ and temperature $T$.  Applied to the magnetoconductance
curves shown in Fig. 1 (a) for different electron densities, the above scaling
procedure yields the data collapse shown in Fig. 1 (b).
  
\vbox{
\vspace{0.2in}
\hbox{
\hspace{-0.1in} 
\epsfxsize 3.3in \epsfbox{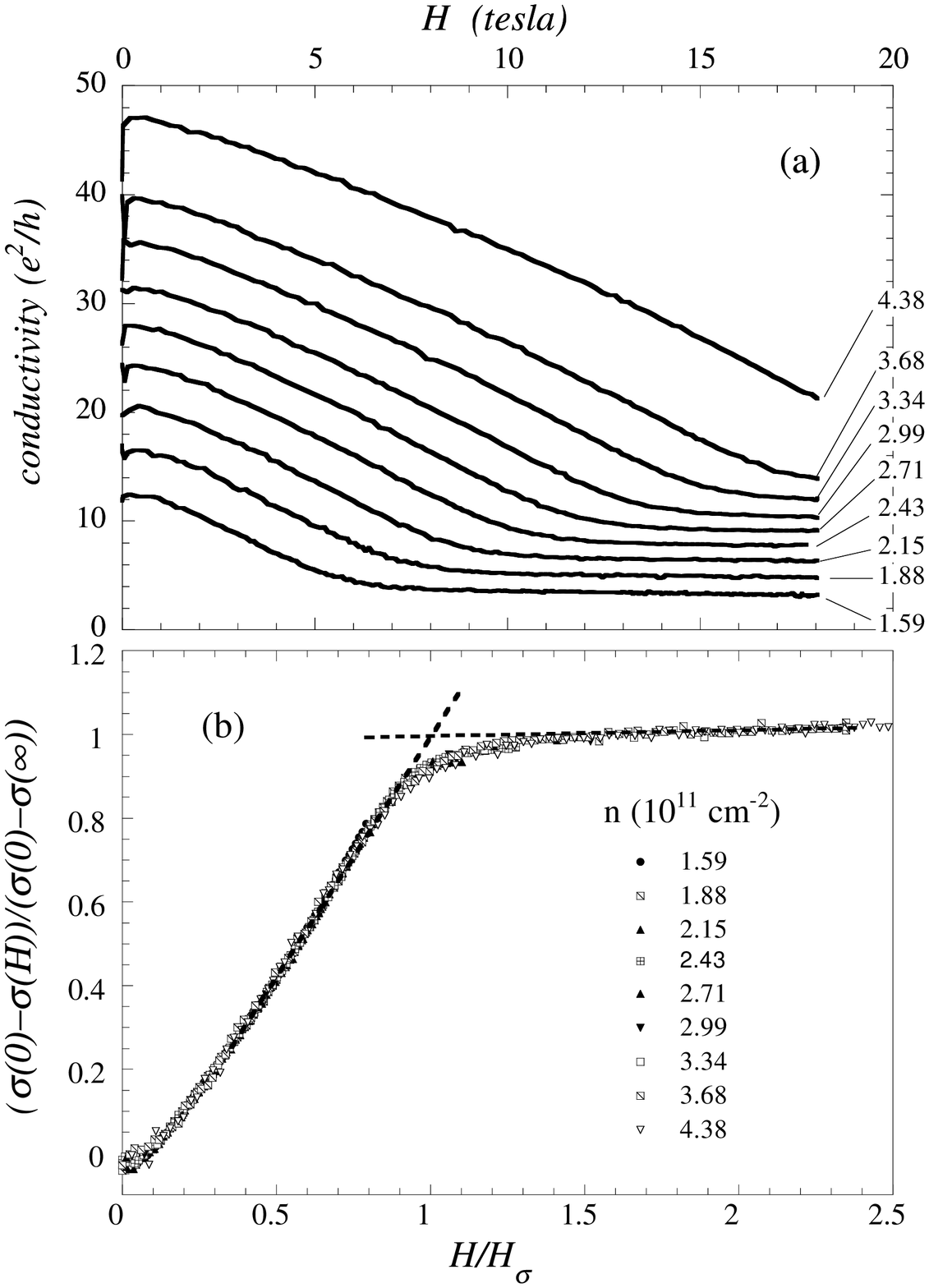} 
}
}
\refstepcounter{figure}
\parbox[b]{3.3in}{\baselineskip=12pt FIG.~\thefigure.
(a)  Conductivity versus in-plane magnetic field at different electron
densities $n$ in units $10^{11}$ cm$^{-2}$, as labeled.  Data are shown
for sample \#3 at $T=100$ mK.  
(b) Data collapse obtained by applying
the scaling procedure, Eq.(1), to the curves of Fig.1 (a).
\vspace{0.0in}
}
\label{1}

Similar scaling holds for curves obtained at different temperatures. This
is demonstrated in Fig. 2, which shows the scaled magnetoconductance for sample
\#2 taken at a fixed density and different temperatures.

There are small departures from a full data collapse with temperature and as the
density is varied.  Possible reasons include: (a) the fact that the
magnetoconductance is not strictly constant at high magnetic field, due possibly
to orbital effects or disorder; (b) interference (weak localization) effects
may become important at very high densities;  (c) one expects scaling to hold
only over a restricted range near a transition.  Nevertheless, within an accuracy
of about 5\%, the magnetoconductance curves for all electron densities and 
temperatures presented here can be mapped onto a universal function using a
single parameter $H/H_{\sigma}$.

\vbox{
\vspace{-0in}
\hbox{
\hspace{-0.2in} 
\epsfxsize 3.3in \epsfbox{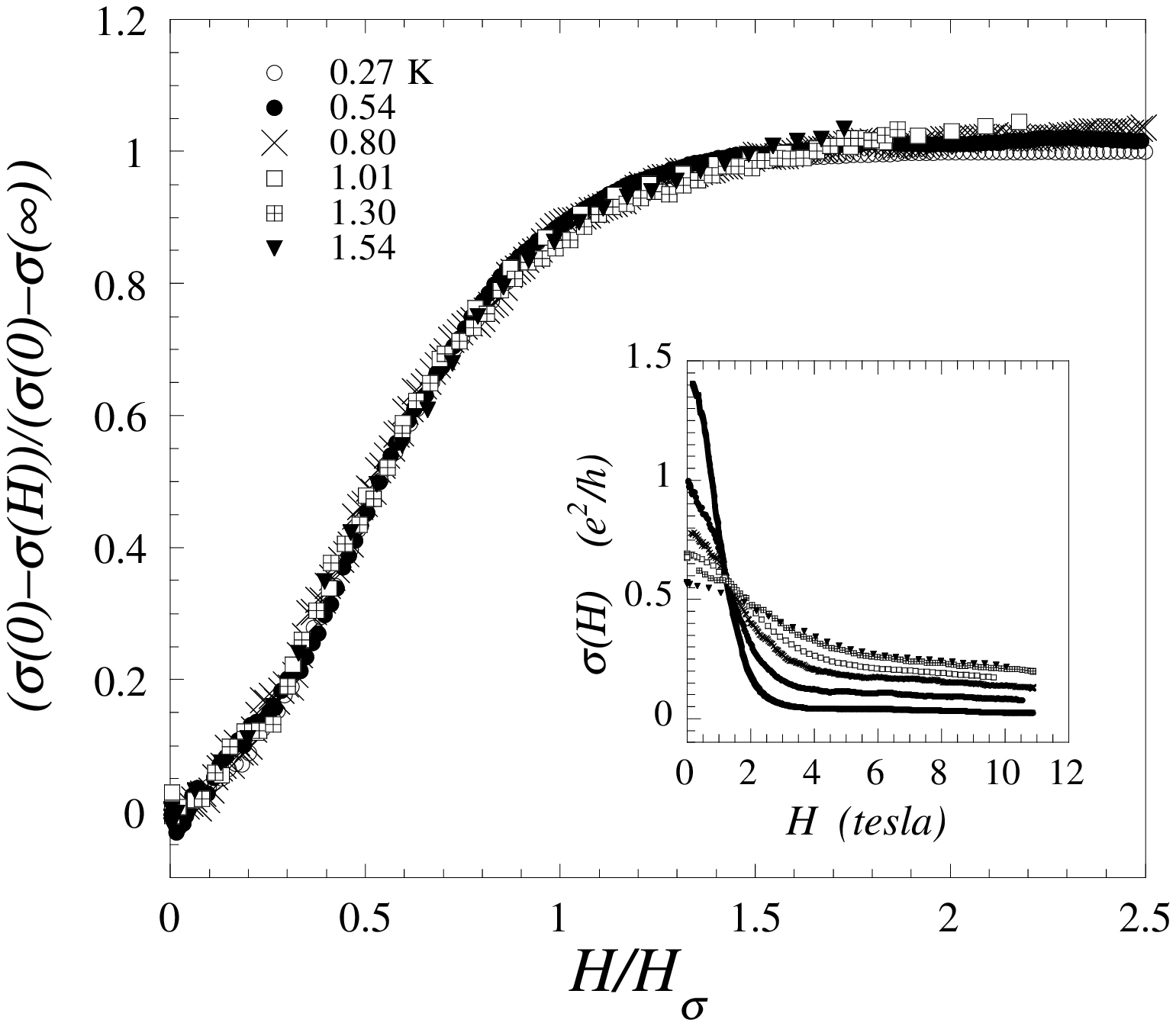} 
}
}
\vspace{-.8in}
\refstepcounter{figure}
\parbox[b]{3.3in}{\baselineskip=12pt FIG.~\thefigure.
Data collapse obtained by applying the scaling procedure, Eq.(1), to the in-plane
magnetoconductivity of the 2D electrons at different temperatures for electron
density $n_s= 0.94 \times 10^{11}$ cm$^{-2}$.  The inset shows the conductivity at different temperatures
as a function of magnetic field.
Data are shown for sample\#2.
\vspace{0.10in}
}
\label{1}

Having thus demonstrated that a one-parameter scaling description of the
magnetoconductance holds for temperatures up to $1.6$ K over a broad range of
electron densities up to  4 $n_c$, we report below the results of
a detailed experimental investigation of the behavior of the scaling parameter,
$H_\sigma$, as a function of temperature and electron density.  In the analysis
that follows, the values of $H_\sigma$ were determined by the intersection of the two dotted lines shown in Fig. 1 (b); this intersection signals a crossover from strong
field-dependence at low fields, $H<H_{\sigma}$, to a conductivity that is almost
independent of magnetic field at $H>H_{\sigma}$.

Fig.3 (a) shows the scaling parameter $H_{\sigma}$ plotted as a function of
temperature for different electron densities above the critical density for the
zero-field metal-insulator transition, $n_c$.  (Data are shown for sample\#$2$,
for which $n_c=0.85 \times 10^{11}$ cm$^{-2}$.)  The scaling parameter becomes
smaller as the electron density is reduced; for a given density,
$H_{\sigma}$ decreases as the temperature decreases and approaches
a value that is independent of  

\vbox{
\vspace{0.in}
\hbox{
\hspace{-0.15in} 
\epsfxsize 3.1in \epsfbox{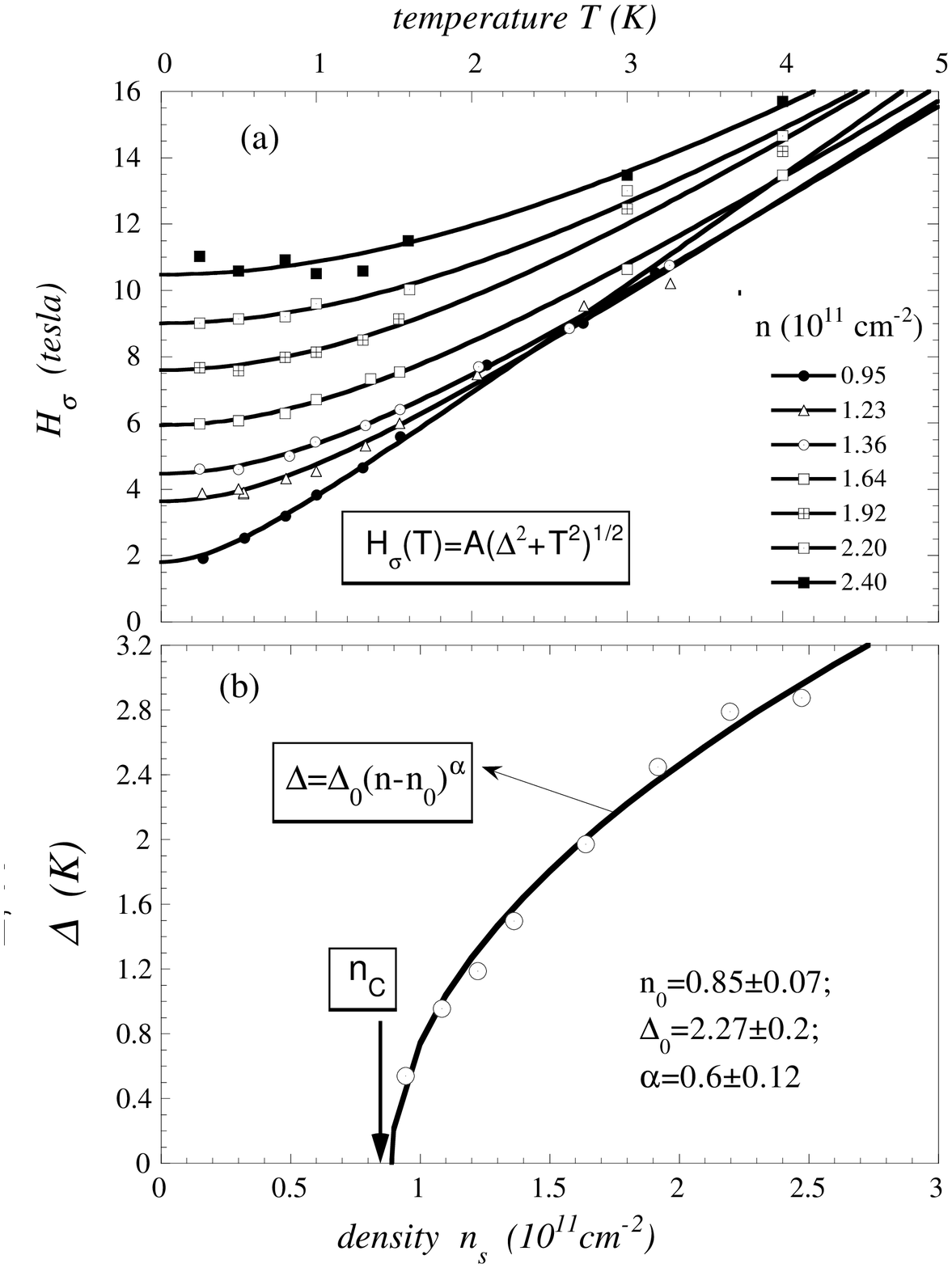} 
}
}
\vspace{.2in}
\refstepcounter{figure}
\parbox[b]{3.3in}{\baselineskip=12pt FIG.~\thefigure.
(a) $H_{\sigma}$ as a function of temperature for different electron
densities; 
the solid lines are fits to Eq.(2).
(b) The parameter $\Delta$ versus electron density; the solid line is
a fit to the expression $\Delta = \Delta_0 (n - n_0)^\alpha$.  Data are shown
for sample \#2.
\vspace{.3in}
}
\label{3}
temperature, $H_\sigma (T=0)$.  As the density is
reduced toward $n_c$, the temperature-dependence of $H_{\sigma}$ becomes stronger
and its low-temperature asymptotic value becomes smaller.  Note that for
electron densities below $1.36 \times 10^{11}$ cm$^{-2}$, $H_{\sigma}$ is
approximately linear with temperature at high $T$.  The behavior of the scaling parameter
$H_{\sigma}(T)$ can be approximated by an empirical fitting function:

$$
H_{\sigma}(n,T)=A (n) [\Delta(n)^2 +T^2]^{1/2} \eqno(2) 
$$

The solid lines in Fig. 3 (a) are fits to this expression using $A(n)$ and
$\Delta (n)$ as fitting parameters.  As can be inferred from the slopes of the
curves of Fig. 3(a), the parameter $A(n)$ is constant over most of the range and
then increases measurably at lower densities (less than $20$\% within the
measured range).  As shown in Fig. 3 (b), the parameter $\Delta$ decreases
with decreasing density and extrapolates to zero at a value $n= n_0$.  Similar
behavior was found for sample\#$1$ (not shown).  Within the experimental
uncertainty of our measurements and of the analysis which yields $\Delta$, a fit
to the critical form, $\Delta= \Delta_0 (n-n_0)^\alpha$, yields a value for
$n_0$ that is within $10$\% of the critical density $n_c \approx 0.85 \times
10^{11}$ cm$^{-2}$ for the metal-insulator transition obtained from zero-field
transport measurements.  Additional measurements are required to determine whether
$n_0$ and $n_c$ are the same or different densities.  

The parameter $\Delta$ represents an energy $k_B \Delta$ and a correlation
time $\tau_H \sim \hbar/k_B\Delta$).  For high densities and low temperature,
$T<\Delta \sim \hbar/\tau_H$, $H_{\sigma}$ is determined by $\Delta$ and the
system is in the zero temperature limit.  Near $n_0$ the measuring temperature
$T>\Delta \sim \hbar/\tau_H$, the field $H_{\sigma}$ is dominated by thermal
effects and is not in the $T=0$ limit.  At $n=n_0$, the energy $k_B \Delta$
vanishes and the correlation time $\tau_H$ diverges; the parameter
$H_\sigma \rightarrow 0$ as $T \rightarrow 0$; the system exhibits critical
behavior, signaling the approach to a new phase in the limit T=0 at a critical
density $n_0$
\cite{Sondhi}.

\vbox{
\vspace{0in}
\hbox{
\hspace{-0.2in} 
\epsfxsize 3.8in \epsfbox{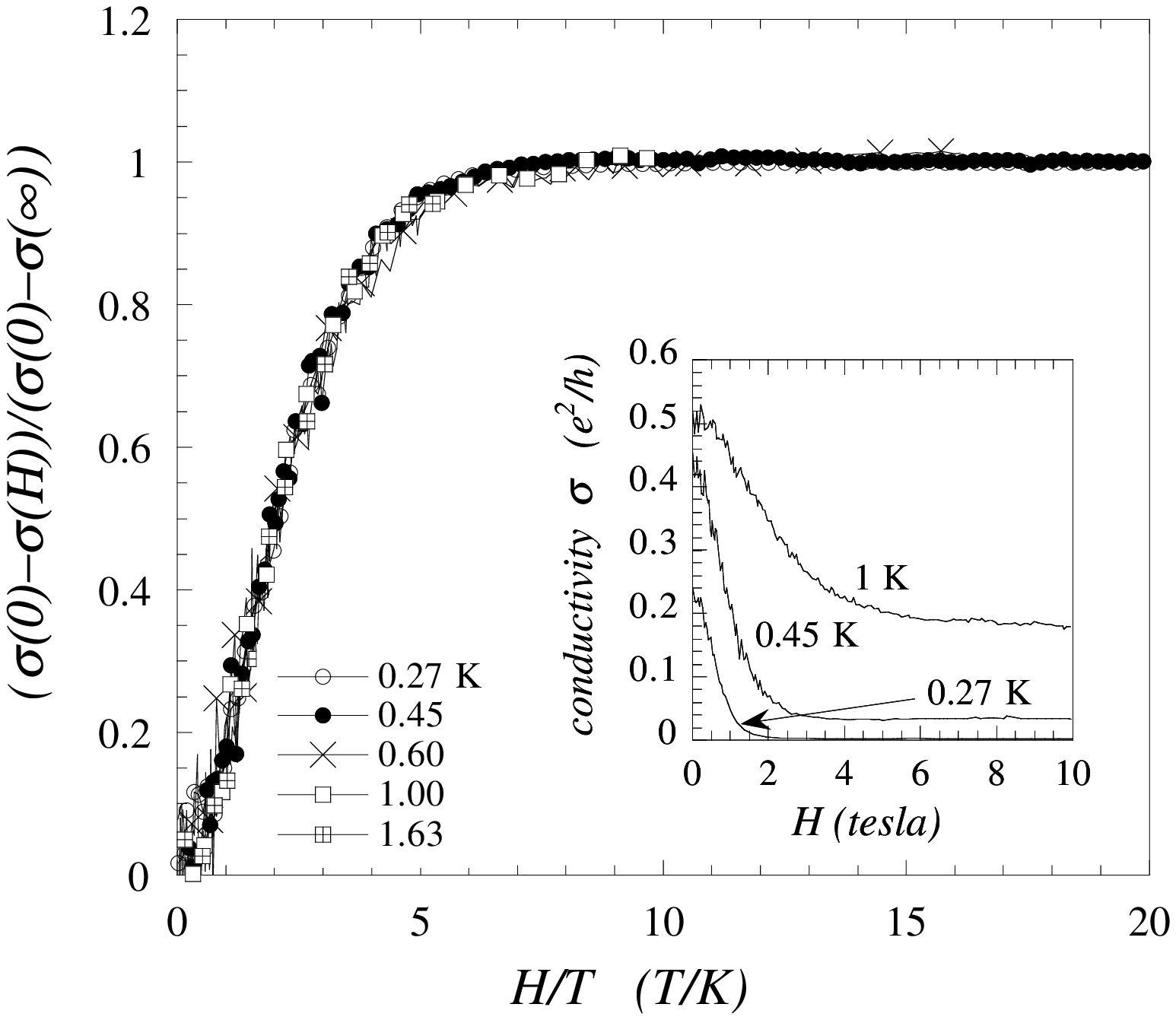} 
}
\vspace{-1.7in}
}
\refstepcounter{figure}
\parbox[b]{3.3in}{\baselineskip=12pt FIG.~\thefigure.
The magnetoconductivity as a function of $H/T$ of sample \#1 at a density $n=0.82
\times 10^{11}$ cm$^{-2}$, just below the critical concentration $n_0$  The
inset shows the conductivity as a function of magnetic field at different
temperatures.
\vspace{0.10in}
}
\label{1}

As discussed earlier, a number of experiments have shown that electron spins play
an important role in the response of dilute 2D systems to an in-plane magnetic
field.  For high electron densities, Shubnikov-de~Haas measurements
\cite{okamoto,vitkalov,vitkalovangular} have shown directly that the unusually
strong field-dependence of the conductivity is associated with the polarization
of the electron spins by the magnetic field.  The fact that the same
scaling, Eq. (1), that applies at these high densities remains valid at low
electron densities implies that spin polarization continues to be a major factor
in the observed magnetoconductivity as $n_0$ is approached.  The parameter
$H_\sigma$, the field that characterizes the system's response to in-plane
magnetic field, vanishes at $n_0$.  Thus, the behavior of the magnetoconductivity
as $\Delta \rightarrow 0$ in the vicinity of $n_0$ indicates critical behavior for
the spin susceptibility.

At $n_0$, $\Delta = 0$, the scaling parameter
$H_\sigma = A T$, and the magnetoconductance scales with $H/T$.  This is
illustrated explicitly in Fig. 4, where the magnetoconductivity of sample \#$1$
is shown as a function of $H/T$ for a density near $n_0$.  
Magnetization that scales with $H/T$ could be associated with 
localized independent spins in a paramagnetic insulator, which would imply 
that the transition at $n_0$ is to a localized phase.
We note, however, that at densities near $n_0$ the resistivity changes from weakly
insulating to metallic temperature dependence, indicating that the localization
length is larger than the average distance between the spins. The spins can
therefore not be considered as independent.

It is possible that the critical behavior we report in this paper
is due to a magnetic instability in the 2D Fermi system at density $n=n_0$. 
Support for this scenario is provided by the unusually strong
temperature-dependence of $H_\sigma$ for metallic densities near $n_0$ and the
increase found in the value of $m^*g^*$ as the electron density is decreased
\cite{okamoto,Pudalov2,Shashkin}, both of which are characteristic features of a
metal near a ferromagnetic instability.  As shown experimentally by Okamoto
{\it et al.} \cite{okamoto} and Vitkalov {\it et al.}
\cite{vitkalov,vitkalovangular}, at high electron densities the ground state of
the system corresponds to complete polarization of the electron spins in high
magnetic fields $H>H_{\sigma}$.  The magnetoconductance continues to scale,
obeying Eq. (1), as the electron density is reduced toward $n_0$, implying there
is no change in the character of the ground state at fields $H>H_{\sigma}$ as
$H_{\sigma}$ decreases.  This suggests that when $H_{\sigma}$ vanishes in the
limit of zero temperature at $n=n_0$, parallel alignment survives at least on a
scale of the localization length $L_{loc}$.  On this basis we suggest that a
transition occurs at $n_0$ due to a ferromagnetic instability.  This could
be associated with long range order or local ordering of ferromagnetic domains or
droplets.

In summary, we have shown that the in-plane magnetoconductivity of the dilute 
2D electron system in silicon MOSFET's can be scaled with a single scaling
parameter $H_\sigma$ over a substantial range of electron densities and
temperatures.  The parameter $H_{\sigma}(n,T)$ that characterizes the response
of the system to a magnetic field obeys an empirical relation $H_{\sigma}=A (n)
[\Delta(n)^2 +T^2]^{1/2}$.  The energy $k_B \Delta$ is found to be to
zero at a critical density $n_0$, indicating  the quantum critical regime of the 2D dilute electron system at $n=n_0$.    
These findings provide strong evidence for a zero-temperature quantum phase
transition at density
$n_0$ near (but not necessarily the same as) the density $n_c$ for the
apparent metal-insulator transition.

We thank B. Altshuler for suggesting the scaling procedure, S. Sachdev and D.
Schmeltzer for  valuable discussions, and S. V. Kravchenko and A. Shashkin for
comments.   S. A. V. thanks A. Larkin, B. Spivak, L. Glazman and A. Chubukov for
useful disscussions.  This work was supported by DOE grant
No.DOE-FG02-84-ER45153.   Partial support was also provided by NSF grant
DMR-9803440.

\end{multicols}

\end{document}